\documentclass[report]{geophysics}

\usepackage{amsmath}
\usepackage{amssymb}
\usepackage{amsfonts}
\usepackage{algorithmic}
\usepackage{algorithm}
\usepackage{pdflscape}
\usepackage{ulem, soul}
\usepackage{color}

\newcommand{\beq}{\begin{equation}}
\newcommand{\eeq}{\end{equation}}
\renewcommand{\hat}{\widetilde}
\renewcommand{\hat}{\widehat}
\newcommand{\bit}{\begin{itemize}}
\newcommand{\eit}{\end{itemize}}
\newcommand{\ben}{\begin{enumerate}}
\newcommand{\een}{\end{enumerate}}

\begin{document}

\title{Full waveform inversion \\ with extrapolated low frequency data}

\renewcommand{\thefootnote}{\fnsymbol{footnote}}

\address{Massachusetts Institute of Technology, 
77 Massachusetts Ave, Cambridge, MA 02138. 
Email: yunyueli@mit.edu, laurent@math.mit.edu\\
}
\author{Yunyue Elita Li and 
Laurent Demanet}

\footer{Geophysics manuscript} \lefthead{Li and Demanet}
\righthead{EFWI}

\maketitle

\begin{abstract}
The availability of low frequency data is an important factor in the success of full waveform inversion (FWI) in the acoustic regime. The low frequencies help determine the kinematically relevant, low-wavenumber components of the velocity model, which are in turn needed to avoid convergence of FWI to spurious local minima. However, acquiring data below 2 or 3 Hz from the field is a challenging and expensive task. In this paper we explore the possibility of \emph{synthesizing the low frequencies computationally} from high-frequency data, and use the resulting prediction of the missing data to seed the frequency sweep of FWI.
\\
\\
As a signal processing problem, bandwidth extension is a very nonlinear and delicate operation. In all but the simplest of scenarios, it can only be expected to lead to plausible recovery of the low frequencies, rather than their accurate reconstruction. Even so, it still requires a high-level interpretation of bandlimited seismic records into individual events, each of which is extrapolable to a lower (or higher) frequency band from the non-dispersive nature of the wave propagation model. We propose to use the phase tracking method for the event separation task. The fidelity of the resulting extrapolation method is typically higher in phase than in amplitude.
\\
\\
To demonstrate the reliability of bandwidth extension in the context of FWI, we first use the low frequencies in the extrapolated band as data substitute, in order to create the low-wavenumber background velocity model, and then switch to recorded data in the available band for the rest of the iterations. The resulting method, EFWI for short, demonstrates surprising robustness to the inaccuracies in the extrapolated low frequency data. With two synthetic examples calibrated so that regular FWI needs to be initialized at $1$ Hz to avoid local minima, we demonstrate that FWI based on an extrapolated $[1, 5]$ Hz band, itself generated from data available in the $[5, 15]$ Hz band, can produce reasonable estimations of the low wavenumber velocity models.

\end{abstract}

\section{Introduction}
Since proposed by \cite{tarantola1984inversion}, full waveform inversion (FWI) has established itself as the default wave-equation-based inversion method for subsurface model building. In contrast to ray-based traveltime tomography (e.g. \cite{woodward2008decade}), FWI includes both phase and amplitude information in the seismograms for elastic parameter estimation. 
When working with reflection data, the model update of FWI is similar to a migration image of the data residual, given the background propagation velocity \cite[]{claerbout1985fundamentals,biondi2004angle}. The accuracy of the resolved model is controlled by the frequency band in the data and the accuracy of the initial (background, macro) model. When this initial macromodel is not sufficiently accurate, the iterative process of FWI gets trapped in undesirable local minima or valleys (\cite{virieux2009overview}). The lack of convexity is intrinsic and owes to the relatively high frequencies of the seismic waveforms.

In spite of an extensive literature on the subject, a convincing solution has yet to emerge for mitigating these convergence issues. So far, the community's efforts can be grouped into three categories. 

In the first category, misfit functions different from least-squares have been proposed to emulate traveltime shifts between the modeled and recorded waveforms. \cite{luo1991wave} proposed to combine traveltime inversion with full waveform inversion, where the misfit is measured over the maximum correlation time lag. \cite{ma2013wave} and \cite{baek2014velocity} proposed variants of FWI augmented with warping functions directly estimated from the data. However, these methods move away from the attractive simple form of the least-squares formulation and require additional data processing steps which are themselves not guaranteed to succeed in complex propagation geometries.

In the second category, additional degrees of freedom are introduced to (attempt to) convexify the waveform inversion in higher dimensions. The extension of velocity with a non-physical dimension was first introduced by \cite{symes1991velocity}. Recently, \cite{shen2004wave,symes2008migration,sun2013waveform} and \cite{biondi2014simultaneous} showed that a velocity extension along the subsurface-offset, planewave ray-parameter, or time lag axis is able to describe the large time shift in wave propagation. \cite{warner2014adaptive} extended the model space to include trace-based Wiener filters comparing the modeled and recorded data. 
\cite{vanLeeuwen2013GJImlm} extended the model space to include the whole wavefield so that the reconstructed wavefield fits the data by design. All these methods rely on an iterative formulation to gradually restrict the extended nonphysical model space to the physical model space, which is often a delicate process without guarantees. In addition, they introduce significant computational cost and memory usage in addition to the already very expensive FWI.

In the third category, the tomographic and migration components in the FWI gradient are separated and enhanced at different stages of the iterations. Based on a nonlinear iterative formulation, \cite{mora1989inversion} decomposed the wavefields according to their propagation directions. Then the low and high wavenumber components of the velocity model are extracted using correlations of these directional wavefields. Following a similar approach, \cite{tang2013tomographically} proposed to enhance the tomographic components at the early iterations and gradually reduce its weights towards convergence. \cite{alkhalifah2015scattering} further separate the gradient components based on the scattering angle at the imaging point. Although these methods do enhance the low wavenumber components of the FWI gradient, the essential difficulty of ensuring correctness of the tomographic component is still mostly untouched.

The most straightforward way to increase the basin of attraction of the least-squares objective function is to seed it with low frequency data only, and slowly enlarge the data bandwidth as the descent iterations progress. However, until several years ago, the low frequency energy below $5$ Hz was often missing due to instrument limitations. More recently, as the importance of the low frequencies became widely recognized by the industry, broadband seismic data with high signal-to-noise ratio between 1.5 Hz and 5 Hz started being acquired at a significantly higher cost than previously. To synthesize some form of low frequency information, \cite{wu2013ultra} proposed using the envelope function. Similarly, \cite{wenyi2014fwi} proposed to use the low frequency information hiding in the difference between data at two adjacent frequencies. Both studies demonstrated that fitting ``manufactured'' low frequency information produces somewhat improved low wavenumber models that enabled closer initialization for the subsequent FWI with the band-limited recordings. However, neither method attempts to approximate the actual low frequency recordings, and neither choice of nonlinear low-frequency combination is justified by a rationale that they succeed at convexifying the FWI objective function. The situation is worse: it is easy to find examples where the wave-equation Hessian of an augmented FWI functional with, say, the envelope function and a lowpass filter, still have undesirable large eigenvalues growing as the frequency increases.

The premise of this paper is that the phase tracking method, proposed in \cite{Li2015Phase}, is a reasonably effective algorithm for extrapolating the low frequency data based on the phases and amplitudes in the observed frequency band. A tracking algorithm is able to separate each seismic record into atomic events, the amplitude and phase functions of which are smooth in both space and frequency. With this explicit parameterization, the user can now fit smooth \emph{non-oscillatory} functions to represent and extrapolate the wave physics to the unrecorded frequency band. Although the resulting extrapolated data can only be expected to accurately reproduce the low frequency recordings in very controlled situations, they are nevertheless adequate substitutes that appear physically plausible in a broad range of scenarios. We are not aware that there is any other attempt at synthesizing low frequency data in the seismic literature. (The mathematical problem of provinding tight guarantees concerning extrapolation of smooth functions from the knowledge of their noisy samples has however been solved in our companion paper \cite[]{demanet2016extrapolation}.)

In this paper, we test the reliability of the extrapolated low frequency data on two numerical examples in the constant-density acoustic regime. In both cases, the low frequencies between 1 Hz and 5 Hz are extrapolated from the recordings at 5 Hz and above. We demonstrate that although the extrapolated low frequencies are sometimes far from exact, the low wavenumber models obtained from the extrapolated low frequencies are often suitable for initialization of FWI at higher frequencies. 

\section{Method}
\subsection{Review of full waveform inversion with truncated Gauss-Newton iterations}
Conventional full waveform inversion (FWI) is formulated in data space via the minimization of the least-squares mismatch between the modeled seismic record $u$ with the observed seismic record $d$,
\begin{equation}\label{eq:fwi-obj}
J(m) = \frac{1}{2}\sum_{r,s,t} (u_s(x_r,t;m)-d_{r,s,t})^2,
\end{equation}
where ${m}$ is the slowness of the pressure wave, $s$ indexes the shots, and $x_r$ are the receiver locations. We also write sampling at the receiver locations with the sampling operator $S$ as $u_s(x_r,t;m) = {S} u_s(x,t;m)$. The modeled wavefield ${u_s(x,t;m)}$ is the solution of a wave equation (discretized via finite differences in both space and time):
\begin{equation}\label{eq:waveeq}
\left( {m}^2\frac{\partial^2}{\partial t^2} - \nabla^2 \right) {u}_s = {f}_s,
\end{equation}
with ${f}_s$ a source wavelet at location $x_s$, and $\nabla^2=\frac{\partial^2}{\partial x^2}+\frac{\partial^2}{\partial y^2}+\frac{\partial^2}{\partial z^2}$ the Laplacian operator.

Starting with an initial model ${m}^{(0)}$, we use a gradient-based iterative scheme to update the model
\begin{equation}\label{eq:grad-updt}
{m}^{(i+1)} = {m}^{(i)} - \alpha \big({\bf J}_r^{\bf{T}}{\bf J}_r\big)^{-1} {\bf J}_{r}^{\bf{T}}{r}({m}^{(i)}),
\end{equation}
where ${m}^{(i)}$ is the model at the $i^{\mbox{\scriptsize th}}$ iteration, ${r} = {S}{u}({m}^{(i)})-{d}$ is the data residual, ${\bf J}_r$ is the Jacobian matrix, and $\alpha$ is the step length for the update. 

In practice, the normal matrix ${\bf J^T_r}{\bf J_r}$ is too large to build explicitly and is often approximated by an identity matrix. In this paper we choose to precondition this matrix by approximately solving the following system using a few iterations of the conjugate gradient method \cite[]{claerbout1985fundamentals,metivier2013full}, 
\begin{equation}\label{eq:gauss-newton}
\min_{\delta m} || {r}({m}^{(i)}) - {\bf J}_r({m}^{(i)})\delta {m}||_2^2,
\end{equation}
where $\delta {m}$ is the unknown increment in model. The linear iteration in (\ref{eq:gauss-newton}) is also known as Least-Squares Reverse Time Migration (LSRTM), which effectively removes the source signature and produces ``true-amplitude'' velocity perturbation at convergence. Due to the limited computational resource and the ill-conditioned matrix ${\bf J^T_r}{\bf J_r}$, we truncate the Gauss-Newton inversion at $3$ iterations for each nonlinear step. Then a careful line search is performed to make sure the objective function (Equation \ref{eq:fwi-obj}) decreases in the nonlinear iterations. (Another effective way of preconditioning the normal operator is to use randomized matrix probing, see \cite{demanet2012probing}.)

To help the iterative inversion avoid local minima, we perform frequency-continuation FWI starting from the lowest available frequency with a growing window. 
Algorithm \ref{algognfwi} shows the workflow for FWI with the truncated Gauss-Newton iterations.

\begin{algorithm}
\caption{FWI with truncated Gauss-Newton iterations}
\label{algognfwi}
\begin{algorithmic}
\vspace{0.5cm}
\STATE Initialize ${m}^0$ 
\FOR{i $\in$ $0 \cdots N $}
\STATE ${r}^{(i)} \leftarrow  {Su(m}^{(i)}) - {d} $
\IF{$||{r}_i || < \epsilon$}
\STATE Converged with model ${m^{(i)}}$
\ELSE
	\STATE Initialize Gauss-Newton iterations
	\STATE ~~~~${p}_0 = {r}^{(i)}$; ${\bf A} = {\bf J}_r({m}^{(i)})$; ${g}_0 = {\bf A}^{\bf{T}} {p}_0$ 
	\STATE First iteration as steepest descent
	\STATE ~~~~${\delta p}_0 = {\bf A}{g}_0$;  ~~~~~~~${\alpha}_0 \leftarrow \frac{\delta {p}_0 ~\cdot~ \delta {p}_0}{\delta {p}_0 ~\cdot ~{p}_0}$
	\STATE ~~~~$\delta {m}_1 = -\alpha_j {p_0}$; ~~~${p}_1 = {p}_0 - \alpha_0 \delta {p}_0$
	\FOR{j $\in$ $1 \cdots n$}
	\STATE ~~ ${g}_j = {\bf A^T} {p}_j$; ${\delta p}_j = {\bf A}{g}_j$
	\STATE
	\STATE ~~ $\left [ \begin{array}{c} \alpha_j \\ \beta_j
	\end{array} \right ] \leftarrow \left [ \begin{array}{cc} 
	 \delta {p}_{j-1} \cdot \delta {p}_{j-1}  &  \delta {p}_{j-1} \cdot \delta {p}_{j} \\
	 \delta {p}_{j} \cdot \delta {p}_{j-1}  & \delta {p}_{j} \cdot \delta {p}_{j} 	
	\end{array}
	\right ]^{-1} \left[ \begin{array}{c}
	{p}_{j} \cdot \delta {p}_{j-1}\\
	{p}_{j} \cdot \delta {p}_{j}
	\end{array}
	\right]$ 
	\STATE
	\STATE ~~ $\delta {p}_{j} \leftarrow \alpha_j \delta {p}_{j-1} + \beta_j \delta {p}_{j}$; ~~~${g}_{j} \leftarrow  \alpha_j {g}_{j-1} + \beta_j {g}_j$ 
	\STATE ~~ ${p}_{j+1}\leftarrow {p}_j - \delta {p}_j$; ~~~~~~~~~~~~$\delta {m}_{j+1}\leftarrow \delta {m}_j - {g}_j$
	\ENDFOR
\STATE Line search to find $\gamma^{(i)} $
\STATE ${m}^{(i+1)}\leftarrow{m}^{(i)} - \gamma^{(i)} \delta {m}_{n+1}$  
\ENDIF
\ENDFOR
\end{algorithmic}
\end{algorithm}

\subsection{Review of phase tracking and frequency extrapolation}
In a previous paper \cite[]{Li2015Phase}, we demonstrated that there exist interesting physical scenarios in which low frequency data can be synthesized from the band-limited field recordings using nonlinear signal processing. This processing step is performed before full waveform inversion in the frequency domain. To extrapolate the data from the recorded frequency band to lower (and higher) frequencies, the phase tracking method consists in solving the following optimization problem to separate the measured data to its atomic event-components: minimize
\begin{eqnarray}\label{eq:objfun-sep}
J_{tracking}( \{ a_j,b_j \}) & = & \frac{1}{2} \, ||\hat{u}(\omega,x)-\hat{d}(\omega,x)||_2^2 \nonumber  \\
    & + & \lambda \sum_j|| \nabla^2_\omega \, b_j(\omega,x) ||_2^2 
    + \mu \sum_j || \nabla_x \, b_j(\omega,x) ||_2^2 \nonumber \\
    & + &  \gamma \sum_j || \nabla_{\omega, x} \, a_j(\omega,x) ||_2^2 ,
\end{eqnarray}
where $\hat{d}$ are the measured data in the frequency domain; $\nabla_k$ and $\nabla^2_k$, with $k = \omega, x$, respectively denote first-order and second-order partial derivatives; $\nabla_{\omega,x}$ denotes the full gradient; and the predicted data record $\hat{u}$ is modeled by the summation of $r$ individual events:
\begin{equation}\label{eq:decomp}
\hat{u}(\omega,x) = \sum_{j=1}^r \hat{v}_j=\sum_{j=1}^r  \hat{w}(\omega) a_j(\omega, x) e^{i b_j(\omega, x)},
\end{equation}
where the wavelet $\hat{w}(\omega)$ is assumed known to a certain level of accuracy. 
The constants $\lambda$, $\mu$, and $\gamma$ are chosen empirically.

The optimization problem for event tracking here is reminiscent of full waveform inversion with high-frequency data, hence shares a similar level of nonconvexity. Yet, by posing it as a data processing problem, the nonconvexity can be empirically overcome with an explicit initialization scheme using Multiple Signal Classification (MUSIC), coupled with a careful trust-region ``expansion and refinement" scheme to track the smooth phase and amplitude. We refer the reader to the detailed algorithm in the previous paper \cite[]{Li2015Phase}.

Having obtained the individual events, we make explicit assumptions about their phase and amplitude functions in order to extrapolate outside of the recorded frequency band. Namely, we assume that the Earth is nondispersive, i.e., the phase is \emph{affine} (constant + linear) in frequency, and the amplitude is to a good approximation constant in frequency -- though both are variable in $x$, of course. A least-squares fit is then performed to find the best constant approximations $a_j(\omega,x) \simeq \alpha_j(x)$, and the best affine approximations $b_j(\omega,x) \simeq \omega \beta_j(x) + \phi_j(x)$, from values of $\omega$ within the useful frequency band. These phase and amplitude approximations can be evaluated at values of $\omega$ outside this band, to yield synthetic flat-spectrum atomic events of the form
\begin{equation}\label{extrap}
\hat{v}^e_j (\omega,x)= \alpha_j(x) e^{i (\omega \beta_j(x) + \phi_j(x))}.
\end{equation}
These synthetic events are multiplied by a band-limited wavelet, and summed up, to create a synthetic dataset. 

The effectiveness of this method for event identification is limited by many factors, chiefly the resolution of the MUSIC algorithm and the signal-to-noise ratio of the data. The algorithm often tracks the strong events and treats the weak events as noise. Moreover, the amplitudes of the events are less predictable than the phases, due to propagation and interfering effects. Therefore, the extrapolated data record is inexact, typically with higher fidelity in phase than in amplitude. In the following section, we test the reliability of the extrapolated low frequencies by initializing the frequency continuation of FWI. Our goal is to bring reliable low wavenumber information in the model by fitting the phase of the extrapolated data and to help enlarge the basin of attraction for FWI when the low frequencies are missing from data. 

\section{Numerical Examples}\label{sec:num}

In this section, we demonstrate the reliability of the extrapolated low frequencies on two synthetic examples. The first model is a wide aperture, one-sided version of the classic Camembert model \cite[]{tarantola1984inversion,gauthier1986two}, where the accuracy of frequency extrapolation can be analyzed carefully on the data record. The second is the Marmousi model, where the frequency extrapolation is tested in a more complex and geologically relevant environment. In both examples, we compare three cases while keeping the initial models fixed: 
\begin{itemize}
\item In the \textit{control} case, frequencies from $1$ Hz to $15$ Hz are used in the frequency continuation of FWI. 
\item In the \textit{extrapolated} FWI case, we first extrapolate the data between $5$ Hz and $15$ Hz to the frequency band between $1$ Hz and $5$ Hz. The extrapolated data are used to build the low wavenumber model to further initialize the frequency continuation starting at $5$ Hz. 
\item In the \textit{missing low-frequency} case, frequency continuation of FWI starts from the lowest frequency at $5$ Hz. 
\end{itemize}


\subsection{Synthetic example: Camembert model}

In this example, we model the synthetic seismic records on the classic Camembert model. We work with the reflection setting where 41 sources and 401 receivers are evenly spaced on the surface. The maximum offset between the sources and receivers is 4 km. A circular low velocity anomaly ($v=1700$ m/s) is placed in a constant velocity background ($v=2000$ m/s). 

Figure \ref{fig:cmbt-invrecord}(a) shows one record between $5$ and $15$ Hz. The tracking algorithm picked up the two strong reflection events from the top and the bottom of the low velocity anomaly. The black dotted line overlaid on Figure \ref{fig:cmbt-invrecord} denotes the inverted arrival time for each event. The inverted two reflection events are shown in Figure \ref{fig:cmbt-invrecord}(b). The tracking algorithm does not perfectly reproduce the input recording, leaving behind the weaker multiple reflection events trailing the second reflection event. Energy from these missing events will appear as phase and amplitude errors at lower frequency when events interfere with each other.

Following the extrapolation strategy in Equation \ref{extrap}, we extend each event from the recorded frequency band to higher and lower frequency bands. Figure \ref{fig:cmbt-phasefunc} shows the comparison between inverted phase function (solid line) and the extrapolated phase function (dashed line). Although the inverted phase function is not perfectly linear with respect to frequency for each event, the extrapolated phase function represents a reasonable approximation to the inverted phase function.

Figure \ref{fig:gather1-model,gather1-extrap} compares the modeled low frequency shot profile (a) with the extrapolated low frequency shot profile (b). The two shot profiles differ from each other in both amplitude and phase. Discrepancies in amplitude are caused by the crude ``constant amplitude'' assumption, while discrepancies in phase are caused by the interference of the unmodeled weak events. Nonetheless, the extrapolated low frequency shot profile is a reasonable approximation of the modeled one for FWI. 

To test the accuracy and reliability of the extrapolated record, we initialize the frequency continuation of FWI using the extrapolated data. Starting from a constant velocity model ($v=2000$ m/s), Figure \ref{fig:vinit-extrap-flo1-fhi5} shows the inversion result using the extrapolated low frequency  ($1$ - $5$ Hz) data in comparison with the inversion result using the modeled low frequency data (Figure \ref{fig:vinit-model-flo1-fhi5}). Although the positive velocity sidelobes are stronger in the extrapolated result, both inversion tests resolve the low wavenumber structure of the velocity model. 

We continue the frequency continuation of FWI starting from the initial models in Figure \ref{fig:vinit-model-flo1-fhi5,vinit-extrap-flo1-fhi5} and obtain the final inversion results in Figure \ref{fig:vfinal-model-flo1-fhi15,vfinal-extrap-flo5-fhi15,vfinal-model-flo5-fhi15}. Figure \ref{fig:vfinal-model-flo1-fhi15} shows the inversion result in the control case when all the frequencies ($1$ to $15$ Hz) are used in the frequency continuation. The inversion successfully delineates the boundary of the low velocity anomaly. Figure \ref{fig:vfinal-extrap-flo5-fhi15} shows the inversion result when low frequency ($1$ to $5$ Hz) energy is missing from the inversion. Thanks to the low wavenumber information extracted from the low frequency extrapolated data, FWI without low frequencies has converged to a meaningful velocity model. In comparison, FWI starting from a constant background velocity model and $5$ Hz data cannot resolve the low wavenumber velocity structure, yielding mispositioned boundaries of the velocity anomaly, as shown in Figure \ref{fig:vfinal-model-flo5-fhi15}. 

One important issue to notice is that the velocity value of the anomaly has not been fully recovered in any of the three cases. This is primarily due to the limited number of iterations that only enable the updates within the available frequency. In the control case, the inversion will resolve the exact model if we let FWI iterate to convergence -- an unrealistic proposition when the data come noisy. Due to the inaccuracies in the extrapolated data, we stop the inversion after 50 iterations to avoid overfitting the unreliable fine scale features in the data. We compare the pseudo-logs at $x=0$ between the true velocity and the inverted velocity in Figure \ref{fig:trace-cmbt-comp}. The true velocity anomaly is $-300$ m/s with respect to the background, whereas the inversion results only recovers half the perturbation with side-lobes of similar opposite perturbations. We find that these inverted results match well with a band-limited version of the true velocity model, where frequencies below $1$ Hz and above $15$ Hz are not present in the model. 

Further tests reveal that the lowest frequency that is needed to recover the circular shape of the Camembert model is $1$ Hz (acquisition and iterative scheme fixed). FWI starting from $2$ Hz will only resolve the boundary of the Camembert model, which might lead to erroneous interpretations. For this simple model, our \emph{extrapolation algorithm continues to yield stable results as we increase the lowest available frequency in the recorded data}. With the highest available frequency fixed at $15$ Hz, the lowest frequency for a reliable extrapolation can be as high as $8$ Hz.
 

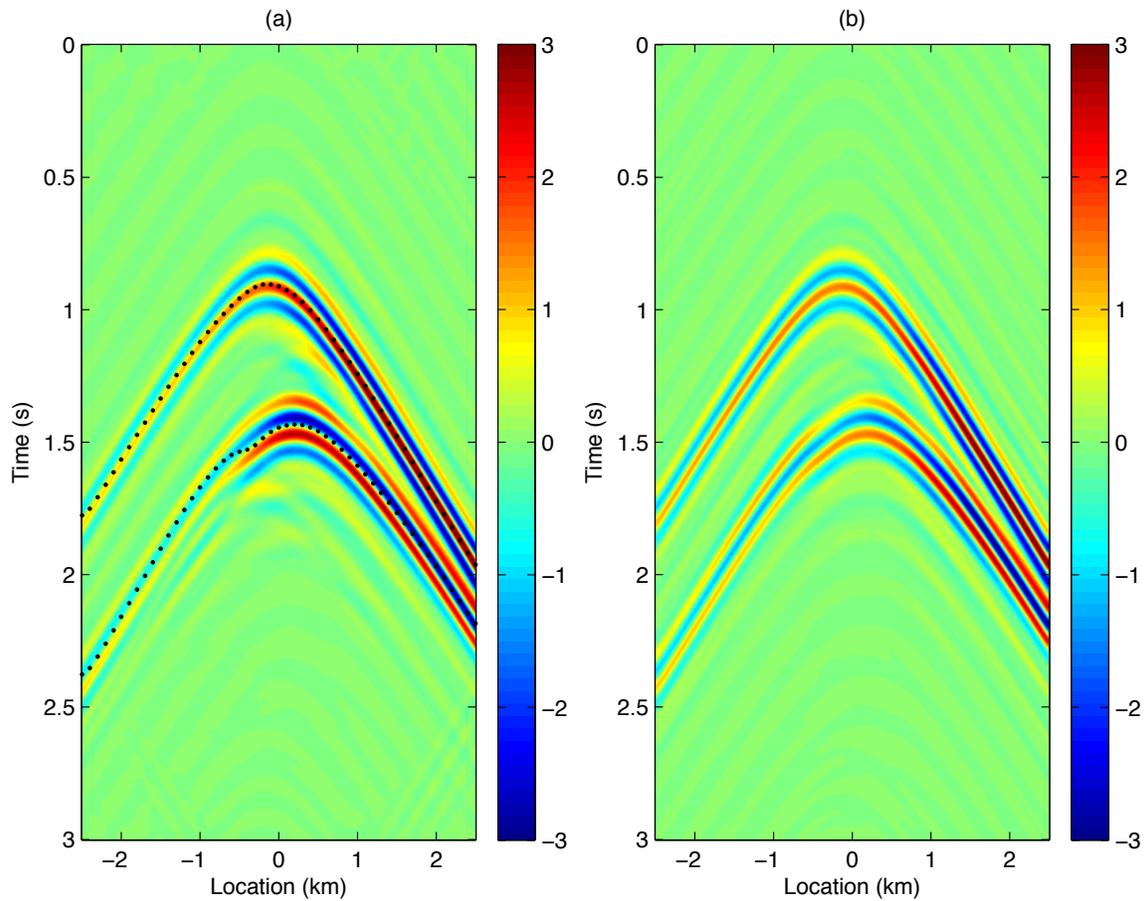
\plot{cmbt-invrecord}{width=\columnwidth}{Comparison between the recorded shot profile (a) and the inverted shot profile (b) on the Camembert model within the recorded bandwidth ($5 - 15$ Hz). The inverted shot profile recovers the phase and the amplitude of the two strong reflection events, whose estimated arrival times are denoted by the dotted line in panel (a). The inversion does not recover the weak multiple reflection events behind the second reflection event. Energy from these missing events will show up as phase and amplitude errors at lower frequency when events become interfering with each other.}
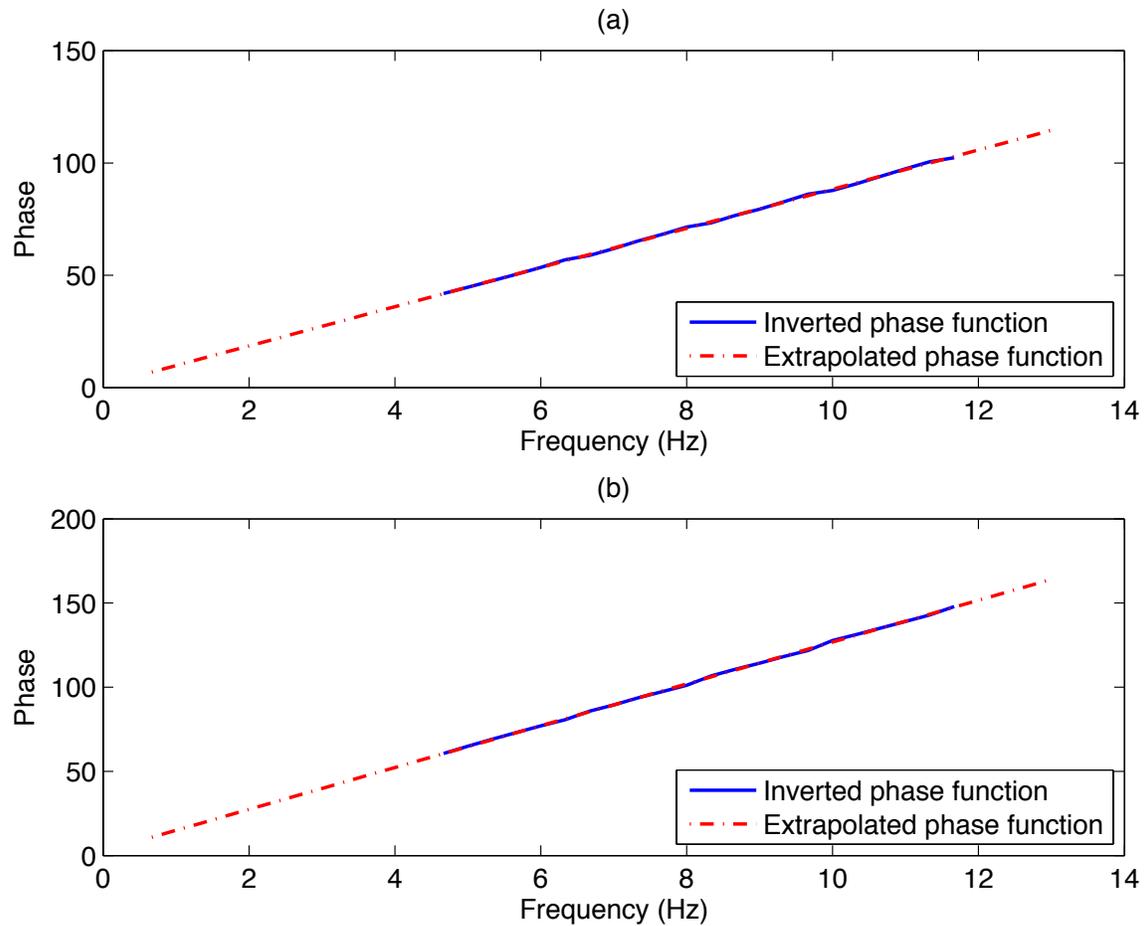
\plot{cmbt-phasefunc}{width=\columnwidth}{Comparison between the inverted phase function (solid line) and the extrapolated phase function (dashed line) for top (a) and bottom (b) reflection events. The extrapolated phase function represents a reasonable approximation to the inverted phase function.}
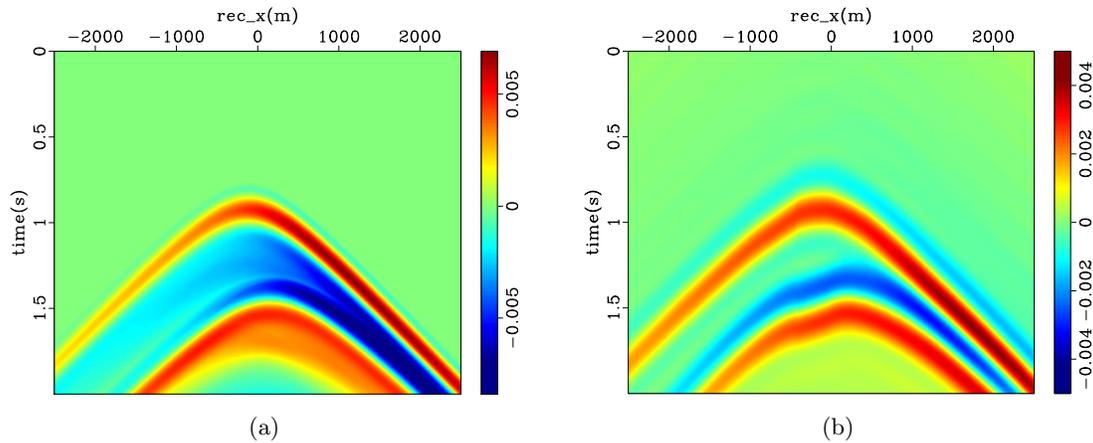
\multiplot{2}{gather1-model,gather1-extrap}{width=0.45\columnwidth}{Comparison between the modeled low frequency ($1$ - $5$ Hz) shot profile (a) and the extrapolated low frequency shot profile (b) on the Camembert model. Discrepancies in amplitude are caused by the crude ``constant amplitude'' assumption. Discrepancies in phase are caused by the interfering of the unmodeled weak events.}
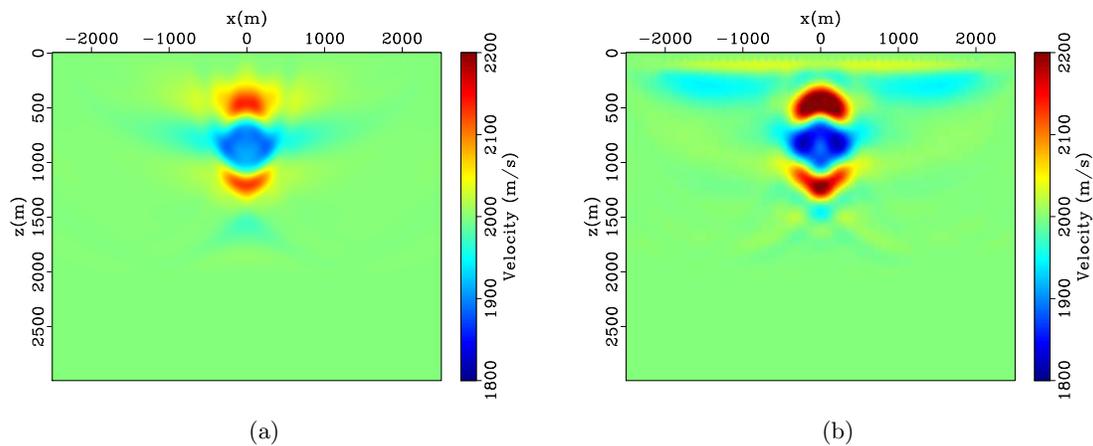
\multiplot{2}{vinit-model-flo1-fhi5,vinit-extrap-flo1-fhi5}{width=0.45\columnwidth}{Inversion results from FWI with the modeled low frequency ($1$ - $5$ Hz) shot profile (a) and the extrapolated low frequency shot profile (b). Both inversion tests resolve similar low wavenumber structure in the velocity model.}

\multiplot{3}{vfinal-model-flo1-fhi15,vfinal-extrap-flo5-fhi15,vfinal-model-flo5-fhi15}{width=0.45\columnwidth}{Comparison between the inverted model from FWI after a full bandwidth continuation. In (a), resulting model from the control case (frequency continuation from $1$ to $15$ Hz). In (b), resulting model from the extrapolated case (initialization using extrapolated low frequencies ($1 - 5$) Hz and frequency continuation with recorded data from $5$ to $15$ Hz). In (c), resulting model from the missing low-frequency case (frequency continuation from $5$ to $15$ Hz).  When all the frequencies ($1$ to $15$ Hz) are used in the frequency continuation, FWI successfully delineates the boundary of the low velocity anomaly. When low frequencies ($1$ to $5$ Hz) are missing from the data, FWI converges to a velocity model with meaningful low-wavenumber components with the aid from the extrapolated low frequency data. Otherwise, FWI cannot resolve the low wavenumber velocity structure, yielding mispositioned velocity boundaries.} 

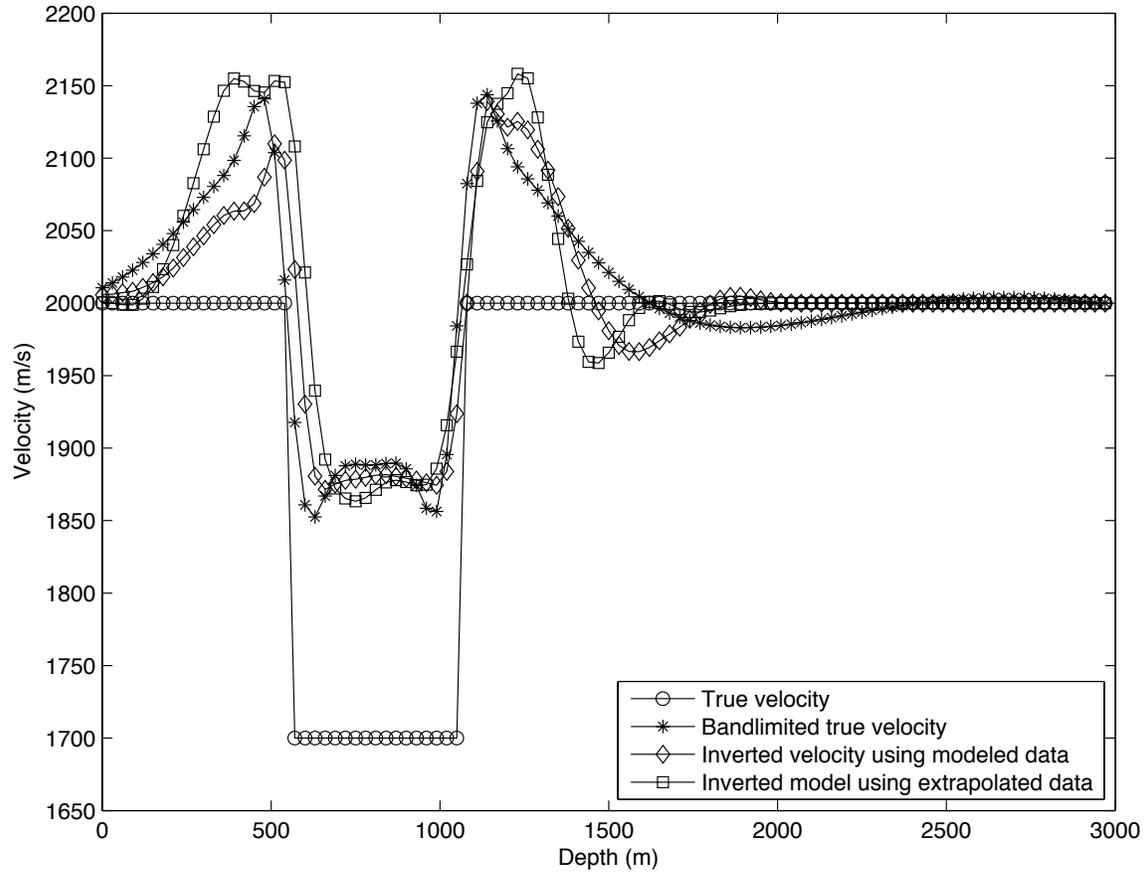
\plot{trace-cmbt-comp}{width=\columnwidth}{Comparison of the pseudo velocity logs at $x=0$ m between the inversion results and the true velocity model. FWI cannot fully resolve the perturbation of the anomaly due to the limited bandwidth in the data. Both inversion results match well with the band-limited (between $1$ and $15$ Hz) true velocity model.}

\subsection{Synthetic example: Marmousi model}

In this example, we test the reliability of the extrapolated low frequency on the Marmousi model (Figure \ref{fig:marm-true}). We restrict FWI to reflection events only by limiting the maximum offset to $500$ m in the inversion. The starting model for FWI is a $1.5$ D linearly increasing velocity profile (Figure \ref{fig:marm-init-linear}). It is an extremely challenging task for FWI to recover large low-wavenumber discrepancies between the initial model and the true model, especially in the deeper section (below 2 km).

Figure \ref{fig:marm-inv-record} shows the comparison between the modeled band-limited ($5-15$ Hz) shot profile (a) with the inverted shot profile by event tracking (b). The tracking algorithm is set up to identify up to $10$ strongest events in each shot record. On this particular shot profile, nine events are identified. The tracking algorithm recovers the input record very well. Figure \ref{fig:marm-low-record} shows the comparison between the modeled and the extrapolated low frequency ($1-5$ Hz) shot profile, amplitude normalized. Despite the amplitude and slight phase discrepancy, the extrapolated shot profile approximates the modeled shot profile sufficiently well in the low frequency band for the purpose of initializing FWI.

Reliability of the extrapolated low frequencies ($1-5$ Hz) are tested with FWI at these low frequencies. Figure \ref{fig:marm-init-model,marm-init-extrap} shows the inverted velocity model using modeled data (a) and using extrapolated data (b). Although not as detailed as Figure \ref{fig:marm-init-model}, the velocity model inverted using the extrapolated data correctly captures the very low wavenumber component of the true model.  These models are used to initialize FWI with data at higher frequencies. 

Figure \ref{fig:marm-inv-model,marm-inv-extrap,marm-inv-nolow} compares the final inverted results after a full bandwidth FWI continuation. In the shallow region (above $2$ km), velocity models resolved in the control case and the extrapolated case are very similar with accurately imaged fine layers and normal faults. Both models have trouble resolving a high resolution and accurate velocity model in the deep region, because reflections from the dipping reflectors and the anticline structure have not been sufficiently recorded due to the limited offset. In comparison, FWI starting at $5$ Hz yields little meaningful information about the subsurface. The inversion failed to update the low wavenumber structure of the velocity model and placed reflectors at wrong positions. 

Figure \ref{fig:trace-marm-comp} compares pseudo velocity logs at three surface locations. Velocity models in both control and extrapolated cases recover the true velocity model very well above $1$ km. Quality of the inverted model degrades with depth. However, both velocity models capture the low wavenumber components of the velocity model. The maximum updates in the deeper section are as high as $1000$ m/s. The huge velocity error prevents FWI starting at $5$ Hz from converging to the true model.

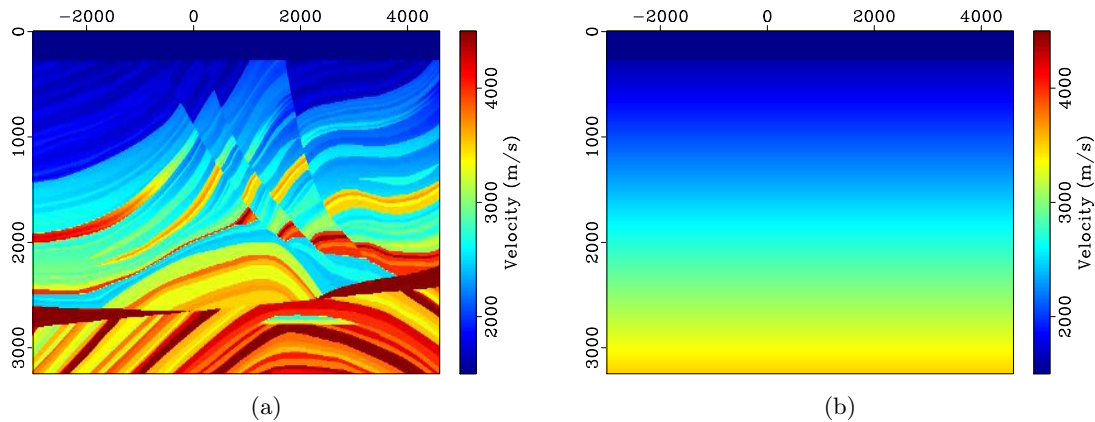
\multiplot{2}{marm-true,marm-init-linear}{width=0.45\columnwidth}{Marmousi model (a) and the starting model for FWI (b). The starting model is a $1.5$ D linearly increasing velocity profile from $1500$ m/s at the water bottom to $3500$ m/s at $3.2$ km. The initial model is far from the true especially in the deeper section.}

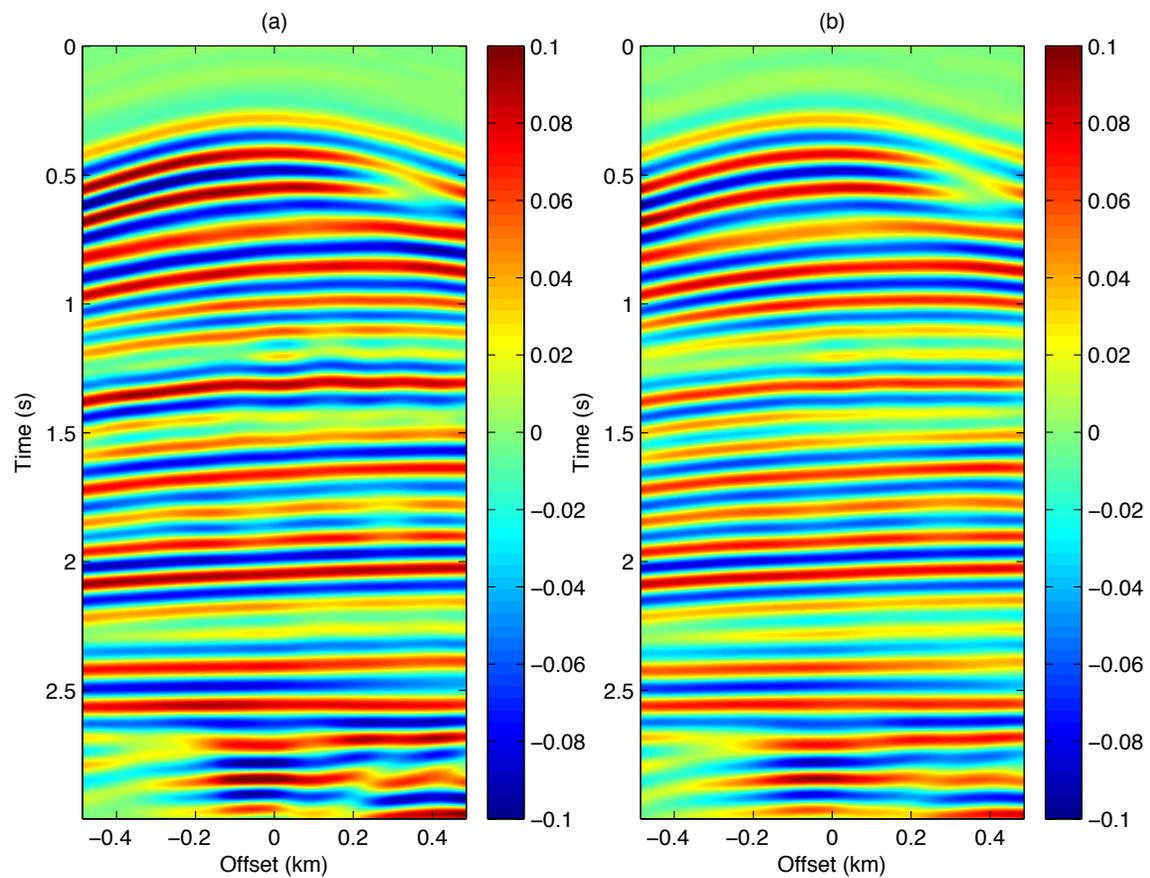
\plot{marm-inv-record}{width=\columnwidth}{Comparison between the modeled band-limited ($5-15$ Hz) shot record (a) and the inverted shot record (b). The tracking algorithm identifies $9$ individual events and recovers the modeled data very well.} 
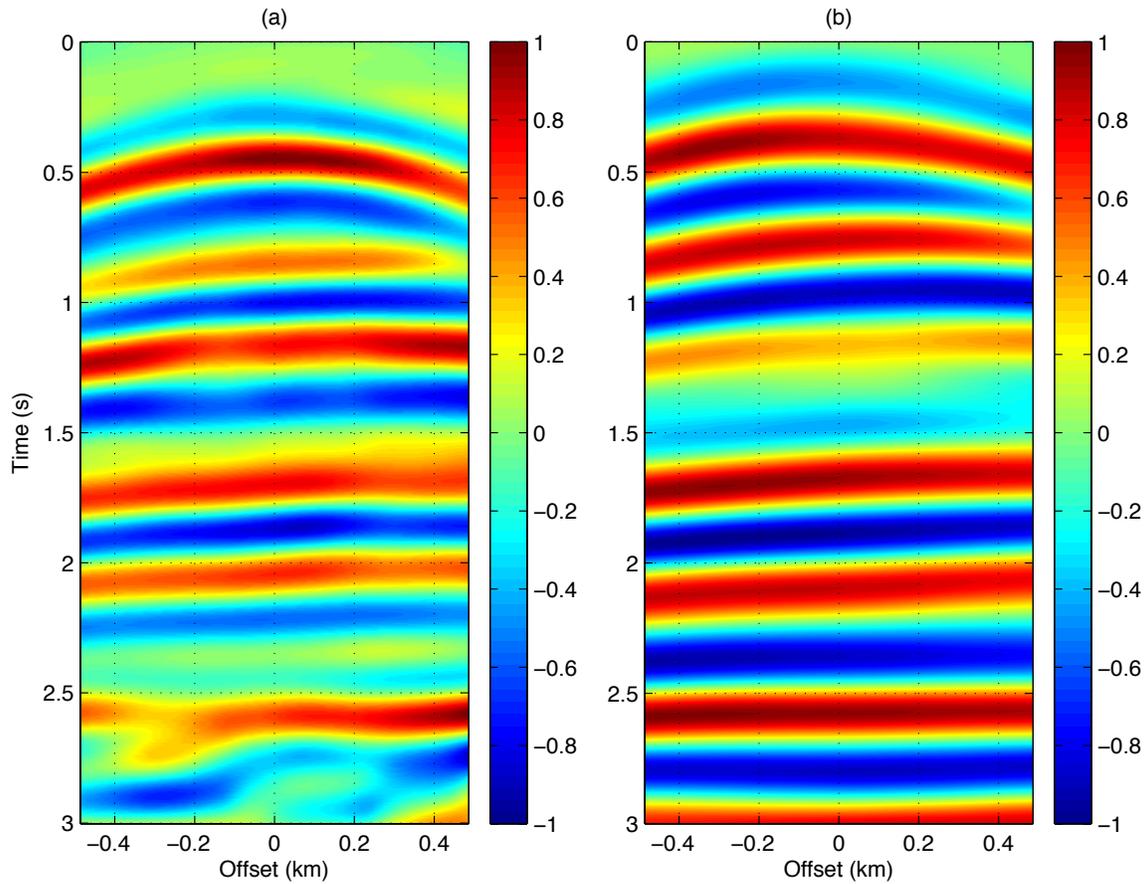
\plot{marm-low-record}{width=\columnwidth}{Comparison between the modeled low frequency ($1-5$ Hz) shot record(a) and the extrapolated low frequency ($1-5$ Hz) shot record (b). The colorbar denotes the normalized amplitude. Despite the imperfect reconstruction of phase and amplitude, the extrapolated low frequencies are adequate for the purpose of initializing FWI.}

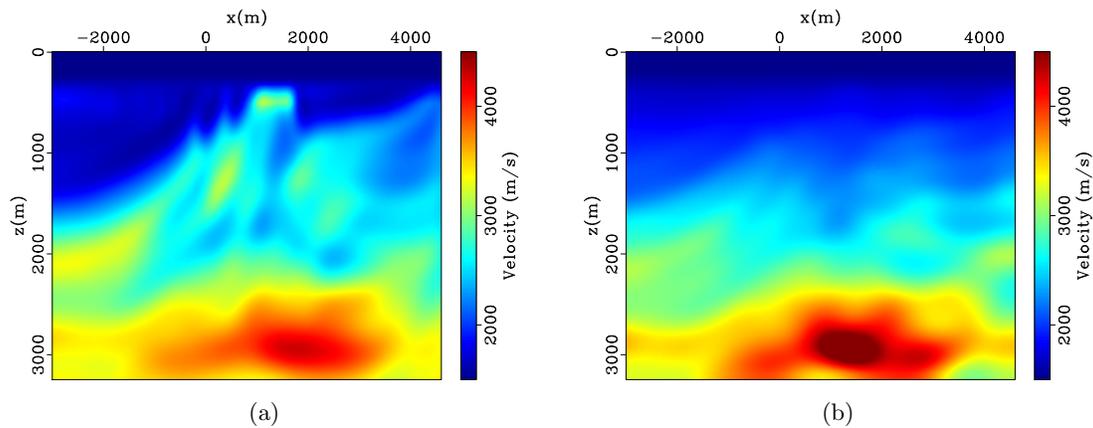
\multiplot{2}{marm-init-model,marm-init-extrap}{width=0.45\columnwidth}{Comparison between the inverted model after FWI using modeled low frequency ($1-5$ Hz) data (a) and using extrapolated low frequency data ($1-5$ Hz) (b). Both models capture the low wavenumber structure of the Marmousi model, although the inverted model using modeled data contains more details in the shallow part.}

\multiplot{3}{marm-inv-model,marm-inv-extrap,marm-inv-nolow}{width=0.45\columnwidth}{Comparison between the inverted model from FWI after a full bandwidth continuation. In (a), resulting model from the control case (frequency continuation from $1$ to $15$ Hz). In (b), resulting model from the extrapolated case (initialization using extrapolated low frequencies ($1 - 5$) Hz and frequency continuation with recorded data from $5$ to $15$ Hz). In (c), resulting model from the missing low-frequency case (frequency continuation from $5$ to $15$ Hz). A better inverted model can be obtained in the control case if we iterate to convergence at the lowest frequencies. However, we limit the number of iterations in the control case to ensure a fair comparison with the extrapolated case.}

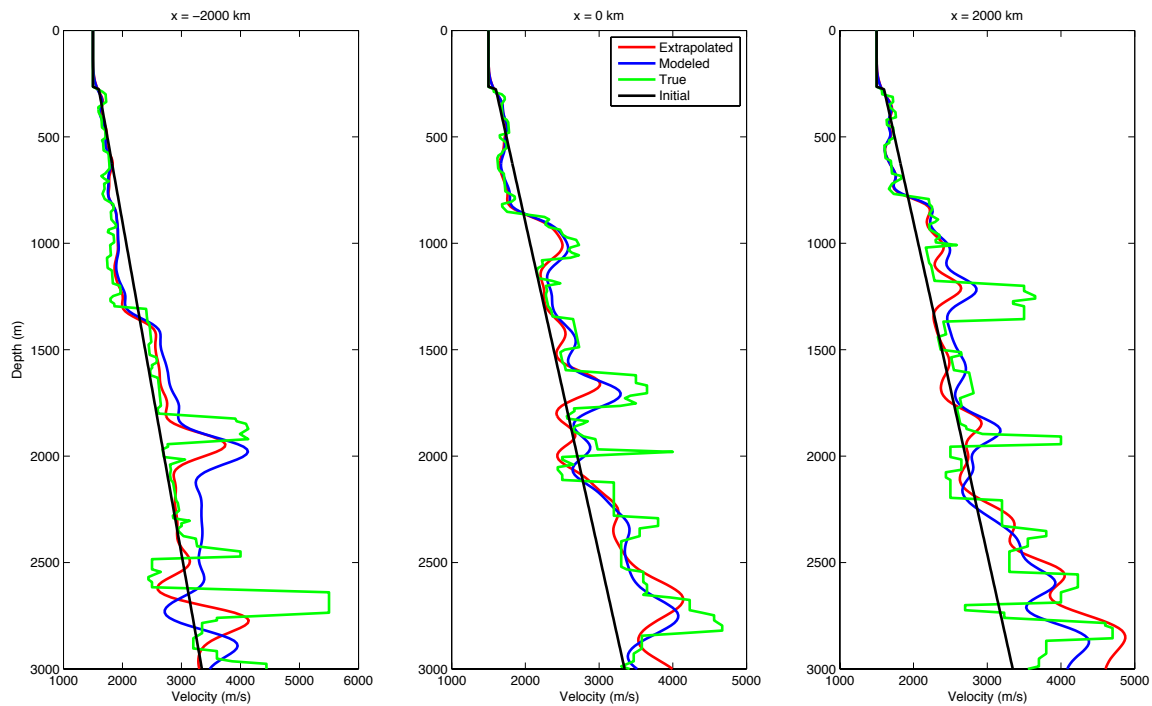
\plot{trace-marm-comp}{width=\columnwidth}{Pseudo velocity logs at three surface locations. The black line denotes the initial model. The green line denotes the true model. The black, green, red, and blue lines denote the pseudo log from the initial, true, control and extrapolated model.}

\section{Discussion}
In the Camembert example, we have carefully examined the accuracy of the phase tracking method for low frequency extrapolation. There are two main reasons that both the amplitude and the phase of the extrapolated data are inexact. First, the tracking algorithm determines the number of individual events as an initialization step. This number may decrease (for event truncation), but may not increase (for even bifurcation) as the tracking expands. These untracked events are the main contributors to the errors in the extrapolated phase function. An aggregation method, with event fragments tracked in subsets of traces and merged into actual composite events, may help improve the accuracy of the event tracking and phase extrapolation.  
Second, compared with the phase extrapolation based on non-attenuative physics, the amplitude extrapolation is less constrained by physical principles. To improve the accuracy of the extrapolated amplitudes, a higher order polynomial or rational function can be used to approximate the amplitude variation with respect to frequency. However, we do not expect the extrapolation to perfectly reconstruct the amplitudes. Consequently, although the extrapolated data are adequate for initializing FWI, they are not suitable for absolute impedance inversion and amplitude-based rock property interpretation.

Due to the inaccuracy in their phase and amplitude, we do not allow FWI to fully fit the data at the extrapolated low frequencies. This limits the resolution of the inverted FWI model. With hundreds of more iterations in the control case, FWI starting at 1 Hz can reduce the data residual to 1\% at each frequency band. Hence,  the velocity model can be fully resolved (with all wavenumber components) because of the availability of both low frequency and long offsets. To the contrary, overfitting the extrapolated data at low frequencies would lead the inversion to undesired local minima and spurious models. FWI with limited number of iterations resolves a good estimate of the velocity model within the increased available frequency band, but it cannot perfectly resolve the model at all wavenumbers due to both the slow convergence and the potential local minima introduced by the inaccuracies in the extrapolated data. 

In our numerical examples, the extrapolated low frequencies are used only to initialize FWI in order to obtain a low wavenumber model. As soon as the frequency continuation moves to the recorded frequency band, the extrapolated low frequencies are abandoned. This leaves the low wavenumber components of the model space unconstrained in later FWI iterations. A proper combination of the extrapolated data and the recorded data needs to be studied to ensure a fully constrained inversion for velocity in the whole wavenumber band.

\section{Conclusion}
To mitigate the nonconvexity of FWI, we propose to start the frequency continuation using the extrapolated low frequency data. The extrapolation is only feasible after decomposing the seismic records into individual atomic events via phase tracking for each isolated arrival. Numerical examples demonstrate that full waveform inversion is surprisingly tolerant to inaccuracies in the amplitude and phase of the extrapolated events. Initializing with the extrapolated low frequencies mitigates the severe nonconvexity that FWI suffers from when only high frequency data are available. By explicitly obtaining the phase and amplitude of each individual event, our method shares an important feature with travel time tomography: its ability to extract kinematic information from high frequencies only. We call the method extrapolated FWI, or EFWI for short.



\section{Acknowledgements}
This project was funded by Total S.A. Laurent Demanet is also grateful to AFOSR, ONR, and NSF for funding. Yunyue Elita Li acknowledges the sponsors of the Energy Resource Laboratory at MIT for funding.

\bibliographystyle{seg}
\bibliography{efwi}

\end{document}